# Improving cold atmospheric pressure plasma efficacy on breast cancer cells control-ability and mortality by using vitamin C and static magnetic field

R. Mehrabifard[1], H. Mehdian[1], K. Hajisharifi[1], E. Amini[2]

[1]Department of Physics and Institute for Plasma Research, Kharazmi University, 49 Dr Mofatteh Avenue, Tehran, Iran

[2]Department of Cellular and Molecular Biology, Faculty of Biological Science, Kharazmi University, 49 Dr Mofatteh Avenue, Tehran, Iran

## Abstract

Since the last decades, there have been numerous reports about the interaction of magnetic field (MF) and cold atmospheric plasma (CAP) with the biological systems, separately. In this manuscript, we have investigated the combined effect of CAP with the static magnetic field (SMF) as an effective method for cancer cells treatment. *MDA-MB-231* breast cancer cells were cultured and treated with CAP in different input power and different exposure time in the presence and absence of the SMF. Vitamin C is used in medium, and cell viability is investigated in the presence and absence of this antioxidant compound. The MTT assay has been employed to measure cell survival, and then *T-test* and one-way *ANOVA* are used to assess the significance level of quantitative data. In order to determine the migration rate of cancer cells, wound healing assay has been carried out. Results show that presence of the SMF and vitamin C as well as increasing the input power has a significant role on the attenuation of the survival and migration rate of the cells. The results of the present investigation will greatly contribute to improve the CAP efficiency in cancer therapy through using the SMF and vitamin C as a complement to conventional CAP therapies.

## Introduction

Cold atmospheric plasma (CAP) has been used in a wide variety of scientific fields due to its remarkable properties. In the last decade, CAPs as a partially ionized gas involving active species (positive/negative ions, radicals), UV radiation, and transient electric field [1–3] which can react chemically and physically with living material has found a great interest in the medicine as a novel non-aggressive treatment. The most prominent usage of plasma in biomedicine is bacterial and fungus decontamination [4–7], sterilization of contaminated tissue [8], dental whitening [9], wound healing, dermatology applications [10] and microorganism inactivation [11]. Now it is believed that the low-temperature active species in the plasma provide these wide ranges of applications through chemically reaction with the bio-medium [12]. In recent years, high side effects and low efficacy of common anti-cancer modalities have been created new insights into novel non-aggressive cancer therapeutic methods. In this field, plasma cytotoxic potential of cancer cells and its selective effect has been received great attention recently. It has been found that radical oxygen species (ROS) and radical nitrogen species (RNS) created by CAP interaction with air are hypothesized to intervene the effects observed in the biological system [13, 14]. These reactive species are known to be employed in interaction with proteins and lipids [15].

The effect of the static magnetic field on the biological system was a topic of interest to biophysics researchers for years. As a most unique feature, SMFs are not easy to shield and can easily penetrate to any biological system [16]. Curative properties of the magnetic field have been seen in various medicine fields including wound healing [17] and bacterial deactivation [21]. Based on the findings of Sadri et al. [19], SMF exposure affects the Mesenchymal stem cells growth cycle, differentiation, and alignment. The therapeutic effect of the magnetic field on cancer cells has been also studied in the last decade [20–22]. We now believed that field intensity and gradient of the field are two major factors that affect the treatment results[23]. On the other hand, vitamin C is known as a crucial nutrient in the human diet, and act as key active substance into biological functions such as detoxifying cell activity [24]. Previous studies indicate that vitamin C has both anti-oxidant and pro-oxidative activities [25]. Ascorbic acid could control level of oxidative stress by ROS and RNS quenching or stimulating which circumstance of cells can be verified vitamin C function. The previous studies have suggested ascorbic acid act as pro-drug for formation of $H_2O_2$ and accomplish as a therapeutic agent against cancer related disease [26].

In this paper, a new noninvasive approach, based on the developing of CAP therapies through using the SMF and vitamin C treatment as complemented methods, has been suggested to improve drastically the breast cancer control-ability and mortality. To examine the proposed methods, cancer cells were treated with CAP+SMF in the presence and absence of vitamin C treatment both for short time (during CAP exposure) and long time (incubation *24h* after CAP treatment in presence of SMF). The migration and survival rate of the cancer cells has been studied after various considered treatments. Results show that presence of the SMF and vitamin C as well as increasing the input power of CAP device has a significant role on the attenuation of the survival and migration rate of the cells, as key parameters in breast cancer cell control-ability and mortality.

## 1. Material and methods

### 1.1. Cold plasma device and SMF structure

The plasma device and its corresponding schematic diagram used in the present report to create the CAP is shown in Fig. 1(a) (b). The device consists of A 1 *mm* diameter steel wire rod as a central electrode connected to a radio frequency (RF) power supply at 13.56 *MHz* via an impedance-matching network, and an electrode wrapped around 5 *mm* glass tube (the glass covered with Teflon stripe) which is grounded. The feeding gas argon was used in this experiment with the flow rate of 473-2900 *sccm*.

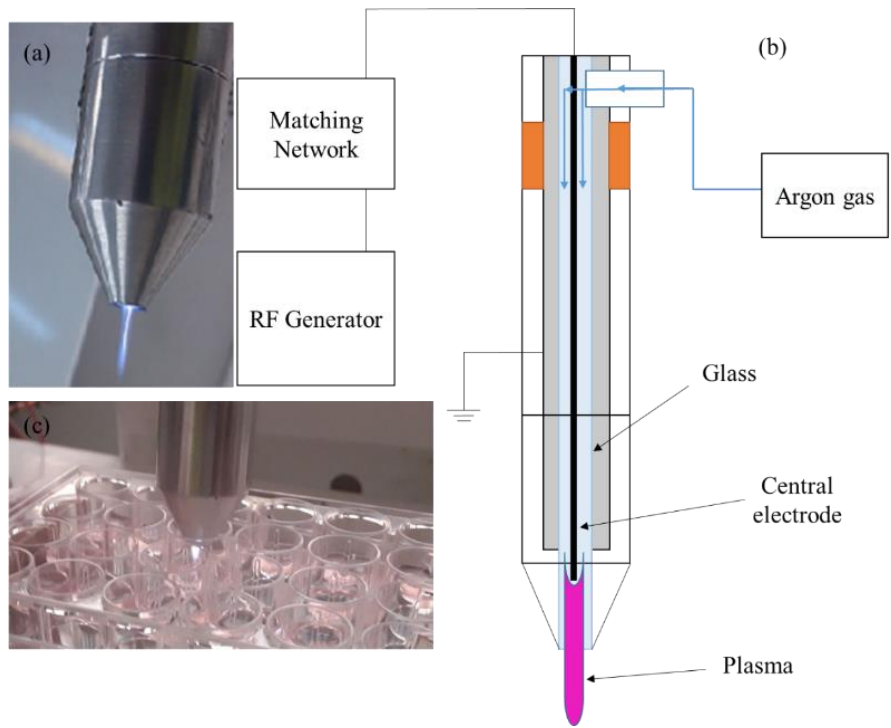

Fig. 1 experimental setup: (a) photograph of the generated plasma, (b) schematic diagram of a CAP device, (c) CAP treatment of the MDA-MB-231 human breast cancer cell SMF

In order to provide SMF, a permanent magnet (IRmagnet, IR) has been used, where the magnitude of MF was measured by Gauss meter (HT201 Portable Digital Gauss Meter). We chose 420 and 270 *mTesla* SMF to test its synergic with plasma on breast cancer cell. *Fig. 2* shows the ways of positioning magnet beneath and side of the cell culture plate. Different parts of the magnet ( $A = C = 470$, $B \approx 0$ ) have been tested in all point as shown in Fig. *2*(b).

## 1.2. Optical emission spectra measurement

In this study the optical emission spectroscopy (OES) as a common technique in plasma diagnostic carried out to determine the element of composition of the atmospheric argon plasma, practically the reactive species. In order to investigate the effect of the magnetic field on the mixing ratio of reactive species in the cold plasma, spectroscopy is carried out for both the presence and the absence of a magnetic field.

## 1.3. Cell culture

The human breast cancer cell line MDA-MB-231 (gift from Pastor) has been cultured in Roswell Park Memorial Institute medium (RPMI-1640) supplemented with 10% Fetal Bovine Serum (FBS) and 1% penicillin and streptomycin. Cultures were maintained at 5% $CO_2$ and 95% humidity at $37°C$. Cells were sub-cultured every 2-3 days when cell confluence reaches 70-85%, and they daily checked under ID03 Zeiss inverted microscope (Carl Zeiss inverted microscope, Germany).

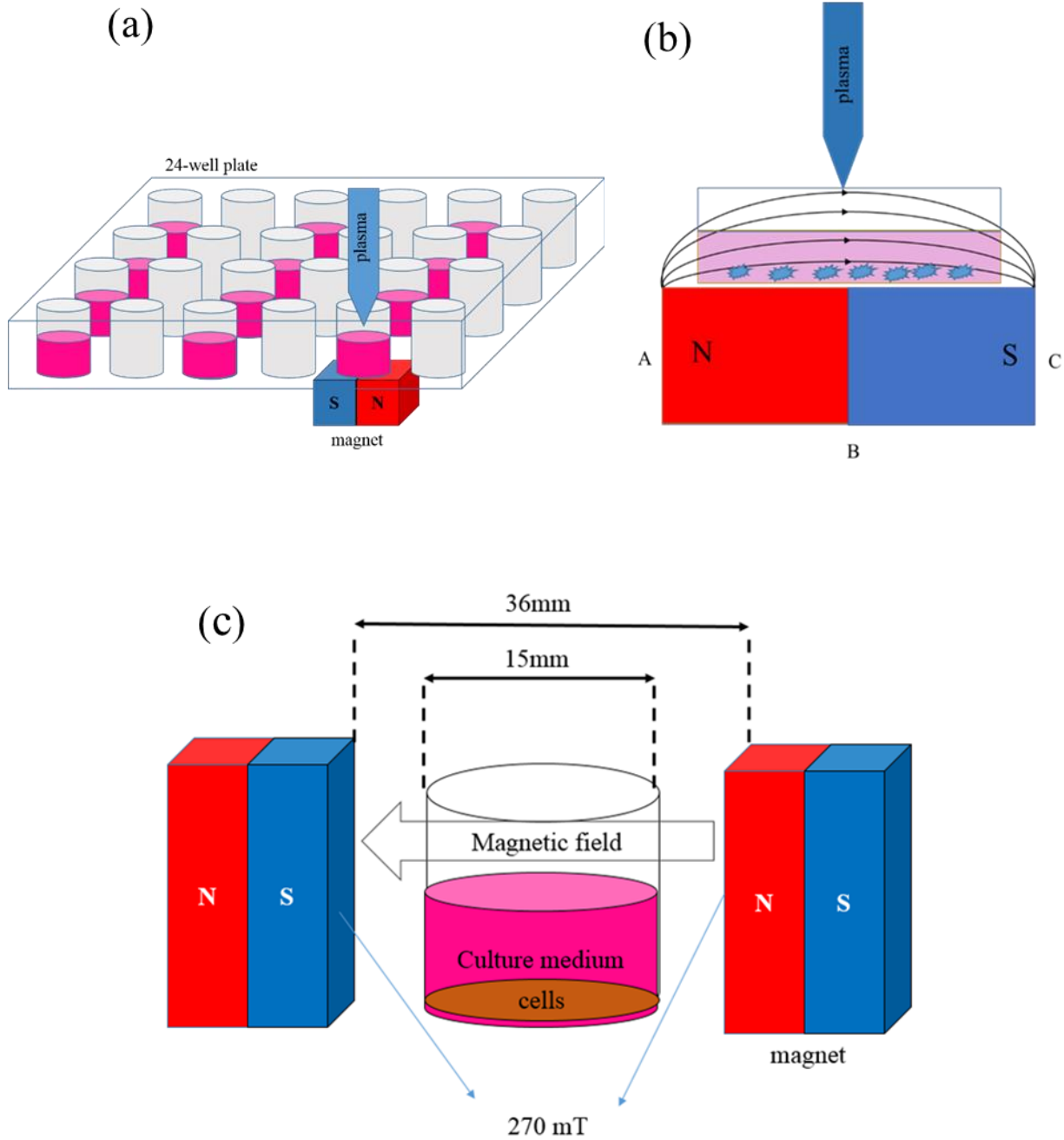

Fig. 2 experiment design: (**a**) MDA-MB-231 cells seeded in 24-well plates and treated with plasma in the presence of SMF, (**b**) cell treated with plasma and SMF at point B, (**c**) presence of SMF after plasma treatment and incubation for 24 h.

### 1.4. CAP and SMF treatment

The plasma treatment was conduct with non-thermal atmospheric pressure plasma as shown in Fig. 1 (c). In order to investigate the synergistic effect of CAP, SMF and vitamin C, the test was carried out in three groups. First, cells were cultured in the 24-well plate at the density of $2\times10^5$ cells per well, subsequently, cells were treated with CAP alone and CAP+SMF in 5, 10, 15, 20 and 25 seconds and 20, 25, 30 and 35 *Watt* input power. As a second group, to investigate the effect of magnetic field alone, the cells were exposed to the SMF for 5-25 seconds. In the last group, the effect of the SMF after plasma treatment (24h incubation in the presence of the magnetic field as shown in Fig. 2 (c)) on the cells viability has been investigated with MTT assay. The vitamin C ($0.2\,mmol\,l^{-1}$) was combined with the cell culture medium two *min* before plasma treatment. CAP+vitamin C effect on cells was investigated in a long time SMF (incubation 24h after CAP treatment).

### 1.5. Cell viability

MTT assay was carried out to estimate the interaction of vitamin C, CAP and SMF with breast cancer cells. MTT assay is a colorimetric assay for evaluating the activity of dehydrogenase enzymes in mitochondria. MTT=3-(4, 5 dimethylthylthiazol-2-yl)-2, 5-diphenyltetrazolium bromide is a yellow tetrazolium salt, which is changed in to purple formazan crystal in living cells. For this purpose after 24h plasma treatment, MTT solution ($5\,mg/ml$) was added into each well, and the plate was incubated for 4 hours at $37°C$. After incubation, the solution was removed and 1 ml *Dimethyl sulfoxide* (DMSO) was added to cells in order to dissolve crystals. The absorption was read by spectrophotometer at the wavelength of $570\,nm$. Cell viability can obtain with equation 1:

$$Cell\ Viability\ (\%) = \frac{treated\ sample\ absorption}{control\ sample\ absorption} \times 100 \qquad (1)$$

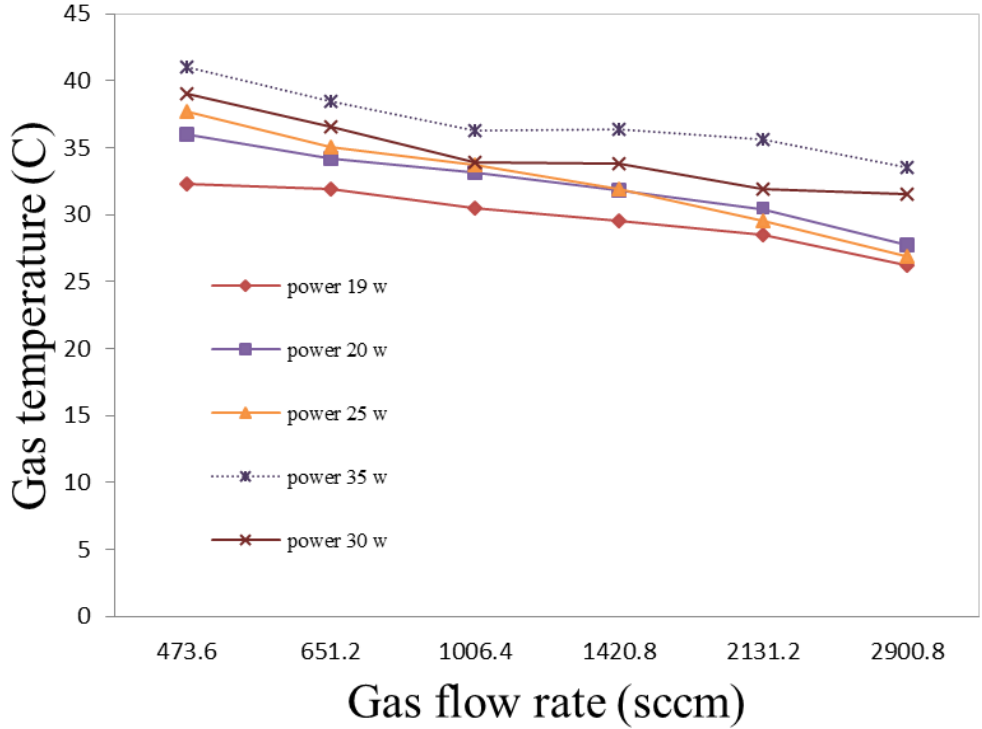

Fig. 3 The dependence of the gas temperature to the gas flow rate in the different input power at the sample point.

### 1.6. Scratch migration assay

To evaluate the effect of CAP, vitamin C and MF on cancer cells metastasis, *MDA-MB-231* breast cancer cells ($2\times10^5$ per-well) were seeded on 6 well-plate. First, the plate coated with $50\mu g/ml$ gelatin and the coated plates were rinsed with PBS and dried. After reaching confluence ($85-90\%$) and scratching by $10\mu l$ pipette tip, cells were exposed to the low power ($20W$) plasma for $5-20s$ exposure time, then at constant input power and time, SMF and vitamin C synergic effect were evaluated. Inverted microscope was used to assess cell mobility into the scratch area.

### 1.7. Statistic

Each experiment was repeated at least three times in triplicate. Statistical analysis has been down to check the statistical significant level (sig was set at $P<0.05$). For comparison between two groups Student *T-test,* and for three and more, one-way ANOVA was applied. Finally, Results were expressed as mean ± SD

## 2. Result and discussion

Motivated by Cheng et al. [27] report, to create the optimal treatment, the SMF+CAP system has been integrated in the way shown in Fig. 2 (a), where the magnet beneath the cell plate has an important role in CAP+SMF treatment at point B. on the other hand, to ensure the prevention of thermal damage of plasma, plasma temperature the position of the cell culture plate (tip of the plasma) has been measured. Fig. 3 shows the gas temperature dependence on the gas flow in different input power. As seen in this figure, for the suitable chosen RF input power as well as gas flow rate, the gas temperature at different applied input power was below $40\,^0C$, which is cold enough to use for biological system. Moreover, the spectrum of the cold argon plasma with and without SMF is shown in Fig. 4 (a) where for the case o presence of SMF shown in Fig. 4 (b), spectra has been measured at spot B. The measurement was done for all input power (but one power is reported). The survey spectra were between 200nm and 1100nm. Referring to the spectra, both spectra are identical in reactive species (RS) production (types of species) but emission intensity increase in presence of SMF.

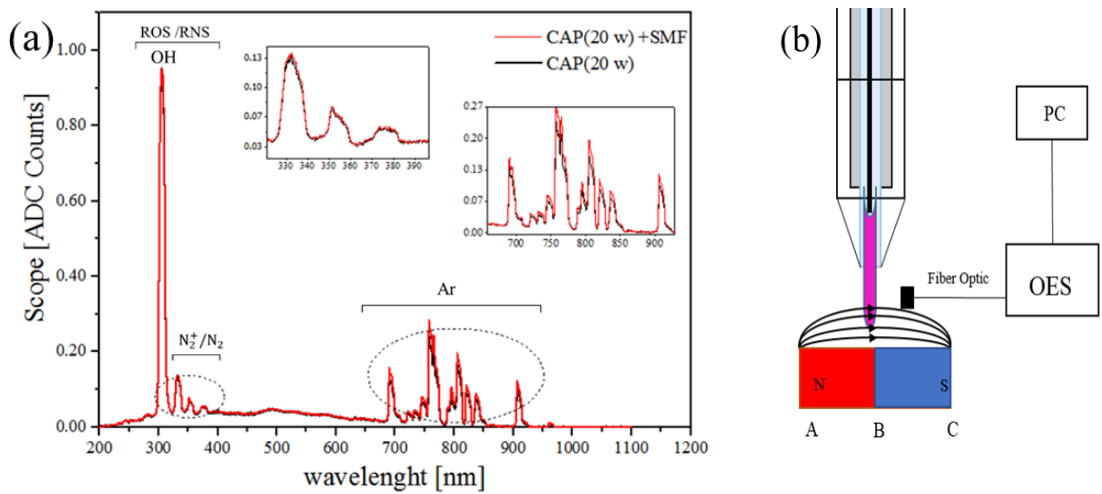

Fig. 4 Optical emission spectroscopy of the CAP and CAP+SMF in the range of 200-1000nm (a). CAP and SMF set up in OES measurement (b)

In order to show magnet effect on the plasma treatment of MDA-MB-231 cancer cells, MTT assay is employed to evaluate the viability of the cells for the plasma treatment in the present and absence of SMF for different value of input power and treatment time, and for the SMF alone in the absence of plasma treatment. Moreover, the student T-test is performed to figure out where the significant lay. Fig. 5 shows that the most of the plasma treatment with SMF had significant level rather than plasma alone. According to the result, the destructive effect of plasma and SMF was augmented with increasing time treatment period and input power. The P-value of 30 and 35w input power for 25s treatment time was lower than 0.01. To prove that the SMF alone cannot induce cell death, cells were placed in SMF for 5-30s. The cell viability was measured 24 hours after SMF treatment. As shown in Fig. 6, there are not significant differences on the cell viability between the five treatments (ANOVA, P-value =0.983).

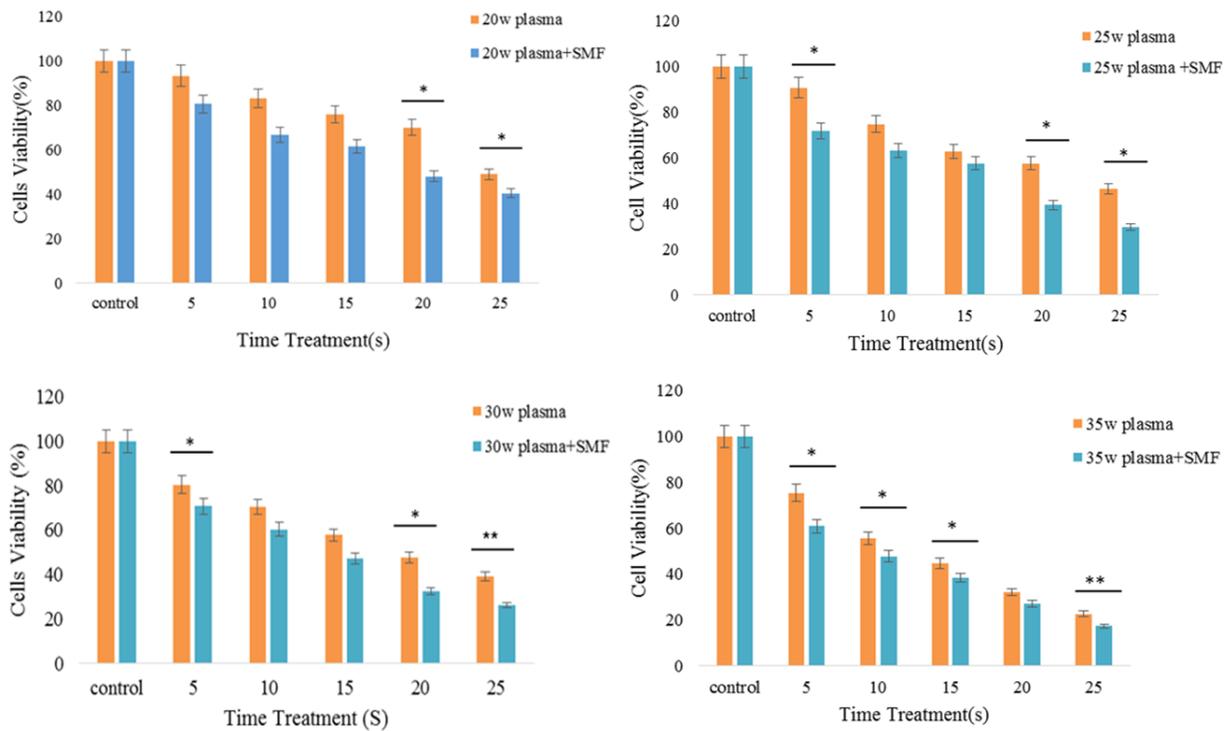

Fig. 5 % viability evaluated from MTT assay on MDA-MB-231 cancer cells in different input power of plasma and 5-25s treatment time with SMF and without it at 24h. (a) 20 w, (b) 25 w, (c) 30, (d) 35w. (P-value<0.05(*), P-value<0.01(**)).

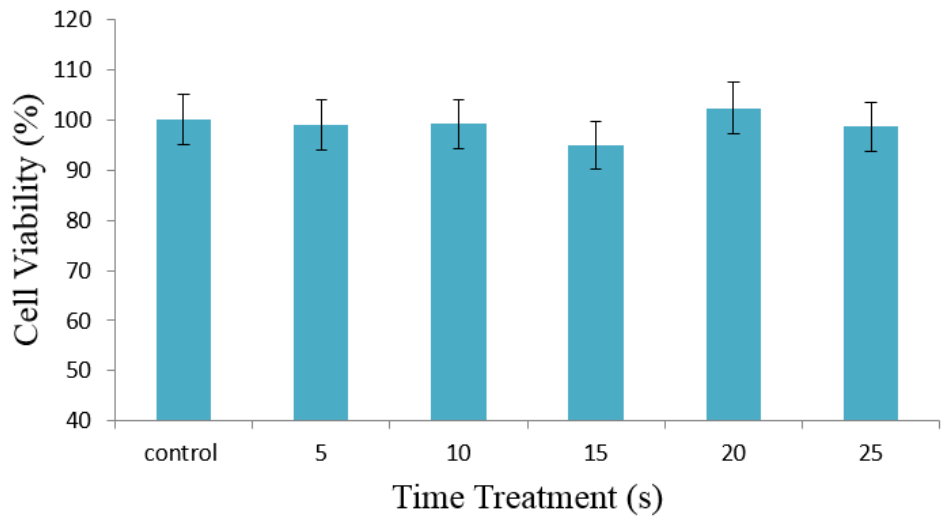

Fig. 6 Cell survival evaluated from MTT assay on MDA-MB-231 cancer cells treated with SMF in 5-25s at 24h.

Another achievement of this study was the evaluation of the combination effect of SMF and cells after plasma treatment on cancer cell survival. Cells were exposed with CAP for 10s and 20s with 25w input power then incubated 24h in the presence of SMF (as a way shown in Fig. 2 (c)). A considerable reduction of cell viability has been observed. As shown in Fig. 7, P-value of student T-test for 10 and 20s CAP treatment is lower than 0.01 and 0.001, respectively.

The effect of the vitamin C and SMF+vitamin C synergetic effect had been investigated in Fig. 8, where the cells viability for constant input power (25w) and constant time treatment 15s has been illustrated. P-value of student T-test showed significance level between groups ($p < 0.001$) and the survival ratio for CAP, CAP+24SMF, CAP+vitamin C, CAP+SMF+vitamin are 68.76%, 22.25%, 62.51%, and 11.45%, respectively. Fig. 9 represents morphologic alternations of cells induced by CAP, CAP+24h SMF, CAP+24h SMF+Vitamin C at 25w input power and 15s treatment time.

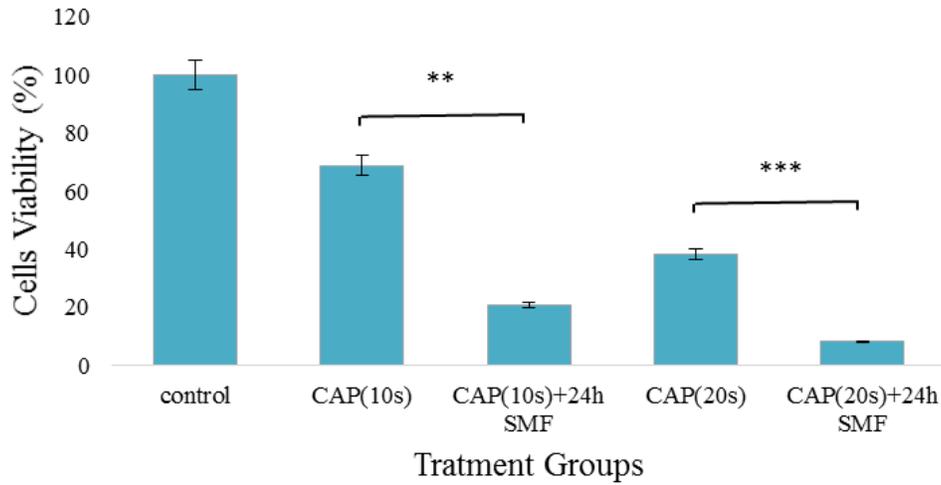

Fig. 7 Cell viability evaluated from MTT assay on MDA-MB-231 cancer cells treated 10 and 20s with CAP and incubated in the presence of SMF for 24h. (P-value<0.05(*), P-value<0.01(**), P-value<0.001(***)).

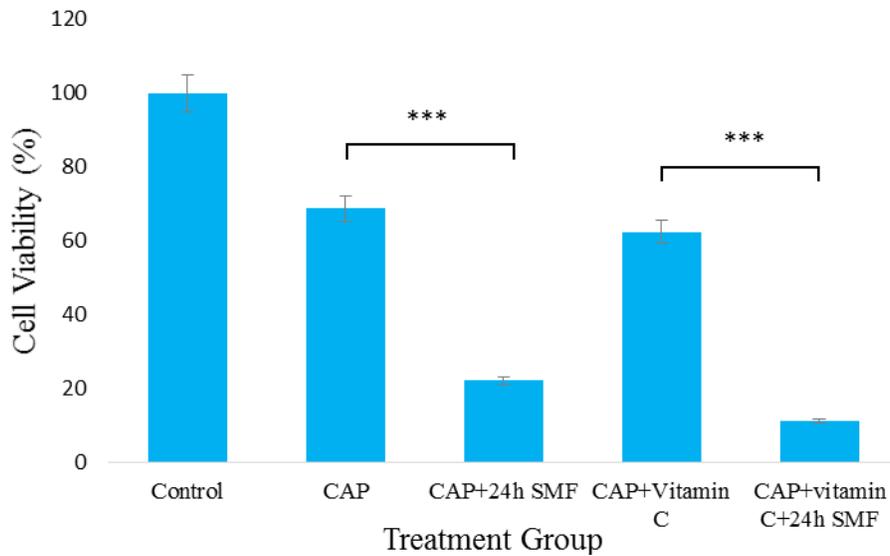

Fig. 8 Cell viability evaluated from MTT assay on MDA-MB-231 cancer cells treated 15s with CAP+24h SMF and CAP+Vitamin C followed by incubated in the presence of SMF for 24h. (P-value<0.05(*), P-value<0.01(**), P-value<0.001(***)).

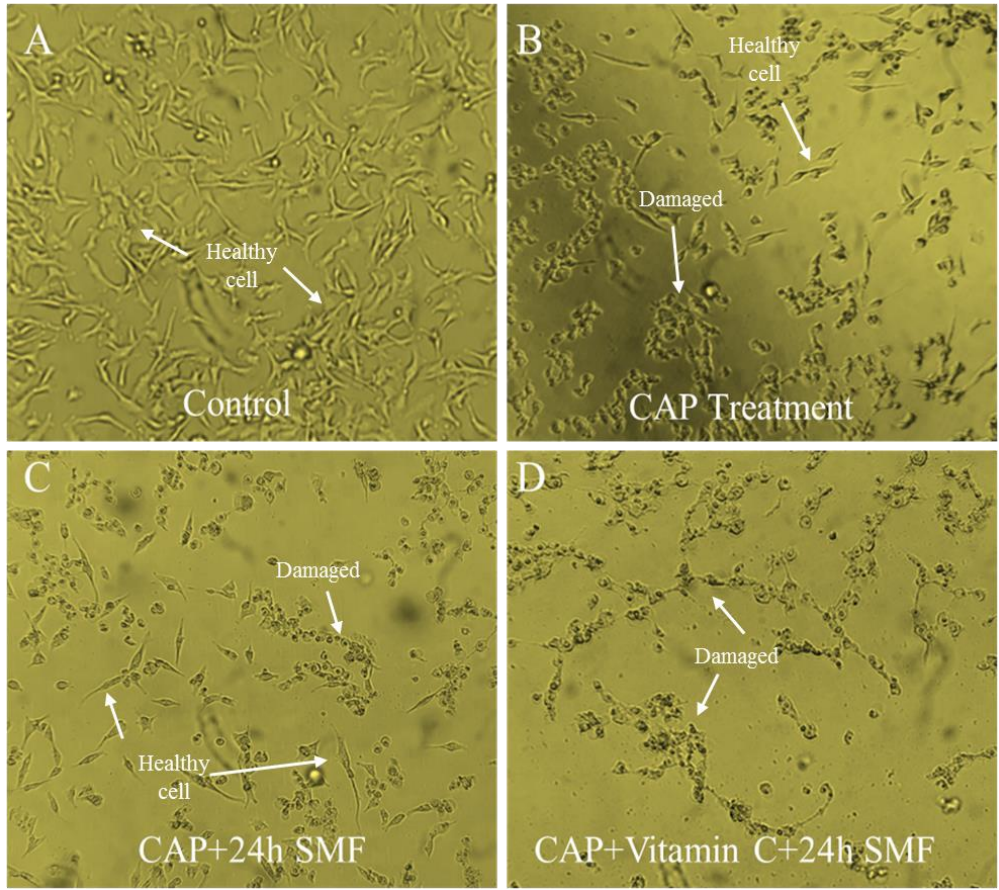

Fig. 9 Morphological change in MDA-MB-231 Cells due to the plasma treatment (A) control (B) CAP treatment (C) CAP+24h SMF (D) CAP+24h+ Vitamin C, constant input power and constant time treatment (15s). Magnification x200.

To more investigation, scratch assay has been performed to clarify whether plasma treatment had modulating effect on breast cancer cells progression as well as cell-cell and cell matrix interaction, specially the adhesion parameter. As shown in Fig. 10, direct plasma treatment notably repressing cell migration activity at constant input power (25W) for 10-25s treatment time after the scratch procedure. With increasing time treatment at constant input power scratch repair process has shown a reduction trend. Also, we investigated the cell migration parameter under various considered treatments including CAP, CAP+ 24h SMF and CAP+24h+vitamin C under constant input power (25w) and time (15s). In all treated groups, cells migration was significantly decreased as compared with the control (Fig. 11). Our results exhibited that this attenuating behavior is higher in the presence of SMF and SMF+vitamin C, which is more important to control the progression of the cancer cell under treatment.

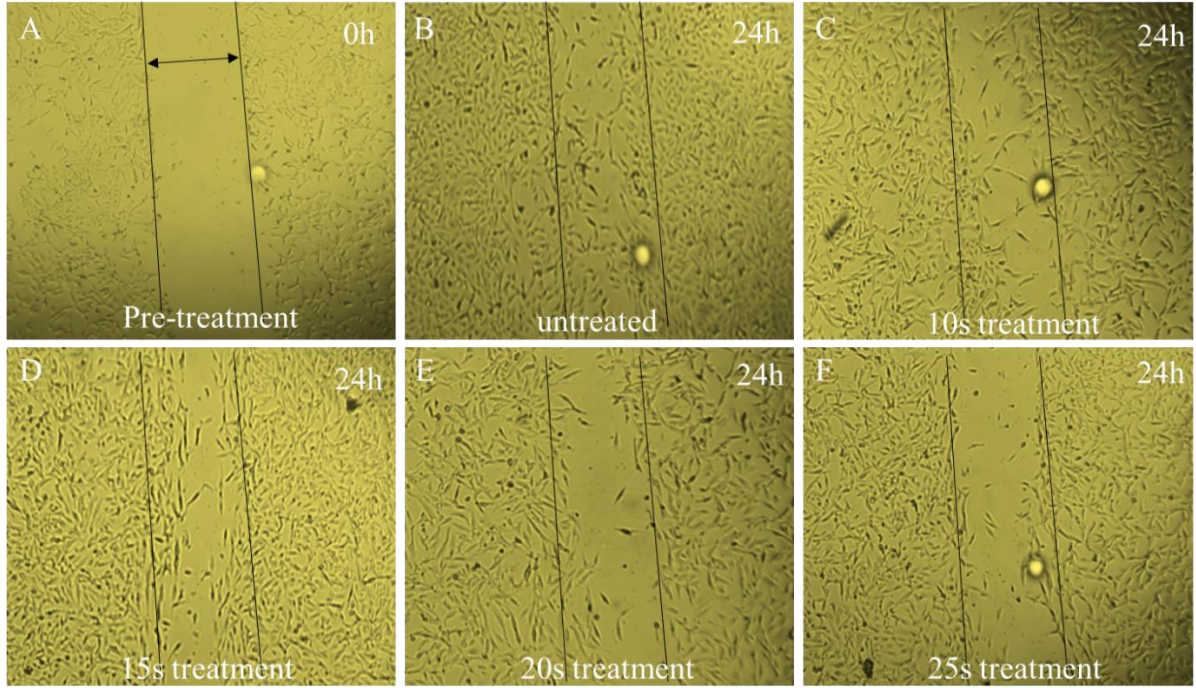

Fig. 10 Plasma treatment decelerated human breast cancer cell migration. MDA-MB-231 cells were grown to confluence and scratched with $10\mu l$ pipette tip. Cell treatment was accomplished with constant input power for 10-25s. Images are taken 24 hours after treatment .Magnification x200.

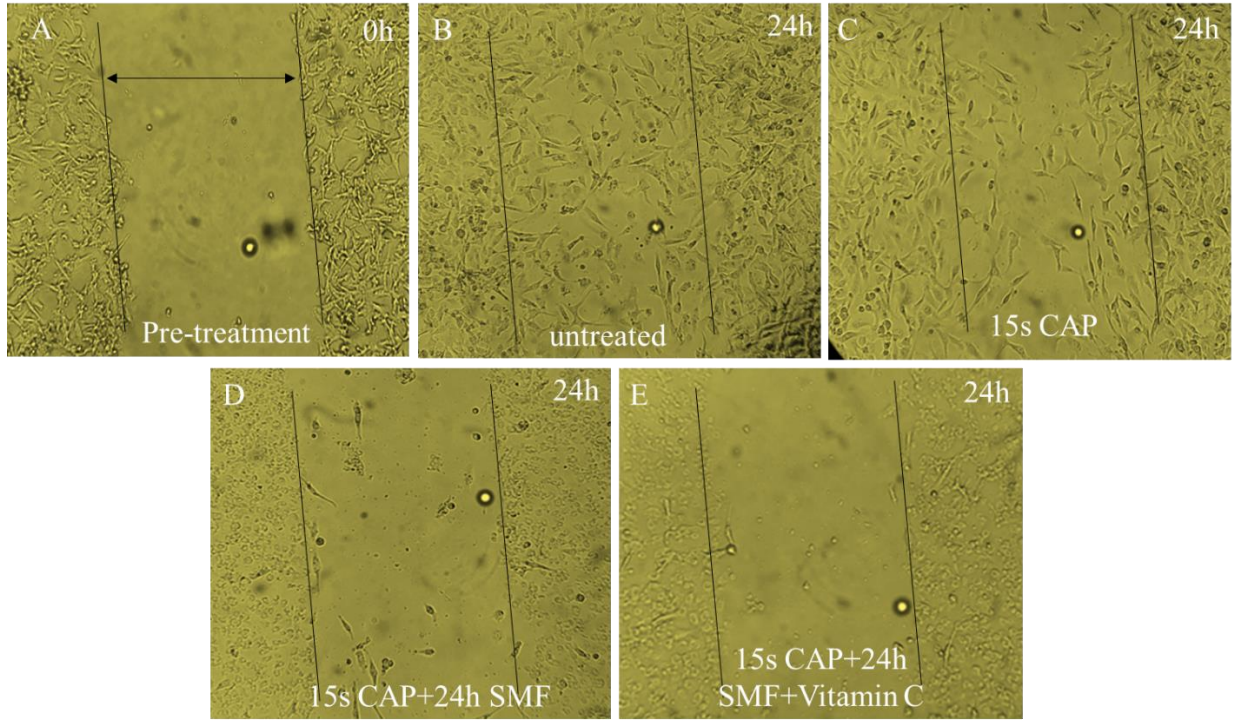

Fig. 11 The effect of CAP + SMF and CAP+ SMF + vitamin C on MDA-MB-231 cells migration. Cell treatment was accomplished with constant input power and 15s treatment time. Images are taken 24 hours after treatment .Magnification x200.

Briefly, we know increasing exposure time could damage normal cells. Therefore, applying effective methods to reduce this side effect are always important in the therapeutic field. The results of this study elucidate that cancer cell survival and rate of migration reduce in the plasma treatment plus SMF exposure and synchronize combination of plasma with SMF and vitamin C rather than plasma alone. The mechanism influence of these processes should be discussed individually.

Nowadays, researches on the plasma treatment medicines indicate that excited $OH$ and $N_2^+$ as reactive species in the plasma plume have a critical roles on the therapeutic effect of CAP in biological systems. Although RS generated in cold atmospheric plasma cannot affect directly the cells, they can initiate a lot of reactions in the liquid phase and then produce a large amount of RS such as ($NO_2\bullet$, $NO\bullet$, $NO_2^-$) and ($NO_3^-$) on it [28–30]. Spectra shows that plasma composition for any arbitrary considered input powers are the same for two cases CAP and CAP+SMF but emission intensity increase in the presence of SMF, which is in agreement with previous result about RS and emission intensity of CAP in the presence of SMF [27, 31]. $H_2O_2$ known as a main species in plasma treated medium that affect the biological system, and it is shown that production of $H_2O_2$ in plasma treated medium is not affected by SMF [27]. Plasma composition constancy indicated that the observed results in CAP+SMF method in contrast with the CAP alone are related to the SMF interaction with cell and RS in the cell culture medium not the amount of RSs.

The most striking result that emerges from the data is the reduction of survival rate of the breast cancer cells that were incubated 24h in the presence of SMF after plasma treatment. This effect was seen when vitamin C was added to the cells culture medium, and the reduction rate increases with increasing treatment time. The further finding confirmed that SMF+CAP method has more effect on cancer cell rather than CAP alone. The possible mechanism of cell death by plasma in SMF could be outcome of two separate reactions: interaction between SMF and cell and interaction between SMF with the CAP-activated medium.

In this regard, numerous researchers worked on the biological effect of SMF and extremely low frequency (ELF) magnetic field alone [32, 33]. Some scientists showed that long-time exposure can inhibit cancer cells growth, and it can have a positive effect on the survival of normal cells [33]. Hakki et al. [34] studied the effect of a low intensity static magnetic field on biological parameters and they showed that decrease in mitochondrial superoxide and OS reduction that could describe the diminish [$Ca^{2+}$]i. The low level of [$Ca^{2+}$]i could explain the increase in membrane potential and the latter the increment in cell growth. Our results exhibited that presence of SMF in different exposure time (5-35s) does not induce any more cell death rather than CAP, but for long-time exposure of SMF (24h) after the plasma treatment the SMF effect on MDA-MB-231 cells is very considerable.

Another experimental aspect of this study was designed on anti-metastatic potential of CAP and SMF. Cell migration is known as an important feature of cancer cells progression [35]. Our results exhibited that CAP treated MDA-MB-231 cell wound areas filled more slowly than untreated after 24h. Examinations show that increasing the treatment time affects the rate of migration. Also, the migration rate was decreased significantly in presence of the SMF and SMF+vitamin C during CAP treatment. According to the results, CAP, CAP+SMF and CAP+SMF+vitamin C may also affect genes related to cell migration and metastasis more than CAP alone. This is in agreement with pervious study on human melanoma cell migration [36].

One of the interesting topics in the last decade has been the treatment of cancer cells by vitamin C [37–39]. J. Young et al. [40] demonstrated that vitamin C can induce apoptosis by ROS generation, attenuation ATP level, and mitochondrial dysfunction. Previous studies have shown that vitamin C prevents cancer cells proliferation and survival, due to the production of $H_2O_2$ [41]. According to these studies, vitamin C is a major factor in the treatment of cancer cells that its synergistic effect with CAP and SMF has been investigated in this study. In addition to production of extracellular ROS ( $NO_3^-, NO_2^-, ONOOH$ ) and RNS ( $\bullet OH, H_2O_2, O_2^-$ ), plasma could generate intracellular ROS signaling that induces apoptotic pathway [42, 43]. ROS enter cells and modulate alterations on intracellular signaling pathway and intracellular components. In this study, we focused on improving plasma efficacy by effective parameter. Therefore, monitoring ROS production on cells is a possible way to explore the mechanism of the combination system

## 3. Conclusion

Although plasma showed selective effect toward the cells, long-time treatment of cells may damage healthy cells. This compels scientists to seek an effective method to treat cancer cells in the shortest possible time. In this study, we presented a new method to improve plasma efficacy by using SMF and vitamin C as complement techniques. we demonstrate that long-time SMF exposure after plasma treatment makes a significant difference in survival rates. Vitamin C was another component of this method that had a positive effect on performance improvement. In addition, scratch assay showed that cold atmospheric plasma plus SMF and vitamin C can be more effective against MDA-MB-231 breast cancer metastasis. Due to importance of novel anti-cancer modalities, the mechanism of these processes should be investigated individually.

In the end, we emphasize that the results of the present investigation will greatly contribute to improve the cold atmospheric plasma therapeutic of breast cancer cells on cell death as well as cancer cell progression where the synchronize treatment including CAP+SMF or CAP+SMF with vitamin C is used instead of CAP alone.